\documentclass[9pt,twocolumn,twoside]{osajnl}

\journal{ao} 

\setboolean{shortarticle}{false}

\usepackage{color}
\usepackage{subfigure}
\usepackage{bm}
\usepackage{cases}

\title{Single-shot non-invasive three-dimensional imaging through scattering media}

\author[1,2,*]{Ryoichi Horisaki}
\author[1]{Yuka Okamoto}
\author[1]{Jun Tanida}

\affil[1]{Department of Information and Physical Sciences, Graduate School of Information Science and Technology, Osaka University, 1-5 Yamadaoka, Suita, Osaka 565-0871, Japan}
\affil[2]{JST, PRESTO, 4-1-8 Honcho, Kawaguchi-shi, Saitama 332-0012, Japan}

\affil[*]{Corresponding author: r.horisaki@ist.osaka-u.ac.jp}

\begin{abstract}
We present a method for single-shot three-dimensional imaging through scattering media with a three-dimensional memory effect.
In the proposed computational process, a captured speckle image is two-dimensionally correlated with different scales, and the object is three-dimensionally recovered with three-dimensional phase retrieval.
Our method was experimentally demonstrated with a lensless setup and was compared with a multi-shot approach used in our previous work~[Y.~Okamoto, \textit{et al}., Opt. Lett. \textbf{44}, 2526--2529 (2019)].
\end{abstract}

\setboolean{displaycopyright}{true}

\begin{document}

\maketitle

\section{Introduction}

Imaging through scattering media has been long studied for biomedical imaging, astronomical imaging, and so on~\cite{bib_merali2015Scatter, bib_gigan2017Scatter, bib_Marx2017AO, bib_Watnik2018Wavefront}.
Recently, optical sensing and control through strongly scattering media, which are difficult to handle in conventional approaches relying on the existence of non-scattered light, have attracted interest in the field of optics and photonics.
The rapidly growing computational power and the improved performance of optical elements for light control drive this area, and various methods have been reported.
These methods are categorized into three types:~feedback-based, inversion-based, and correlation-based~\cite{bib_mosk2012Scatter, bib_horstmeyer2015Scatter}.

The feedback-based approach utilizes wavefront shaping behind or inside scattering media with an iterative feedback process based on an optimization algorithm~\cite{bib_vellekoop2007Focusing, bib_aulbach2011Scatter, bib_katz2011Scatter, bib_katz2012Scatter, bib_vellekoop2010Foucising, bib_park2013Nanoparticle, bib_conkey2012Scatter, bib_Conkey2012ColorScatter, bib_vellekoop2015Scatter}.
Issues with the feedback-based approach are the large number of feedback loops and the need for a probing process to measure the focusing state in the every loop.
The inversion-based approach, including time reversal and phase conjugation, senses and controls the optical distribution through scattering media by taking the inverse of a transmission matrix expressing the scattering process~\cite{bib_yaqoob2008Scatter, bib_popoff2010Scatter, bib_popoff2010Opaque, bib_popoff2011Scatter, bib_Lai2012Scatter, bib_chaigne2013Scatter, bib_benjamin2013Scatter, bib_liutkus2014CS, bib_mounaix2016Scatter, bib_horisaki2016Scatter, bib_edrei2016ScatterDeconv, bib_Horisaki2017Scatter, bib_Horisaki2017SR, bib_Singh2017ScatterPlate, bib_Antipa2018Diffuser}.
These methods realize single-shot imaging and focusing without any feedback process.
However, they need to probe the whole or part of the transmission matrix before the imaging and focusing stage.

The correlation-based approach exploits the shift invariance of speckles, which is called the memory effect~\cite{bib_Feng1988Scatter, bib_Freund1990Scatter, bib_bertolotti2012Scatter, bib_katz2014Speckle, bib_singh2014Diffuse, bib_schott2015Scatter, bib_edrei2016Scatter, bib_li2017ScatterEncryption, bib_Yang2018Scatter}.
In the correlation-based methods, an autocorrelation process is used for removing speckles and exposing object signals in captured images.
An advantage of the correlation-based approach is the lack of a need for the probing process, which is a  drawback of the previous two approaches.
As a result, the correlation-based approach has realized non-invasive imaging through scattering media.
This approach has been recently extended to a three-dimensional case~\cite{bib_takasaki2014Scatter, bib_singh2017Scatter, bib_Shi2017Scatter, bib_Okamoto2019Scatter}.

Drawbacks with the methods for correlation-based three-dimensional imaging through scattering media include the need for capturing multiple images and/or invasive optical processes.
Thus, they are difficult to apply to imaging of dynamical and practical scenes.
Here we present a method for tomographic reconstruction of a three-dimensional object from a single speckle image captured through scattering media without any invasive or probing process. We demonstrated the method experimentally with a lensless setup.
Our method enhances the possibility and practicability of imaging though scattering media for a wide range of applications, including biomedicine, security, and industry.

\section{Method}

\begin{figure}[]
\begin{center}
		\includegraphics[scale=0.7]{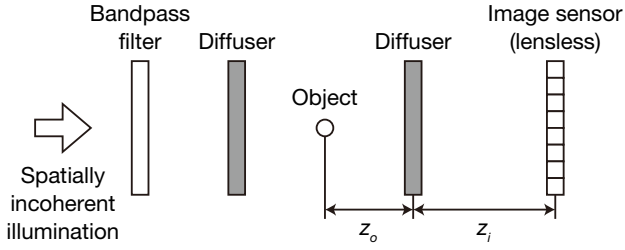}
\end{center}
\caption{Schematic diagram of single-shot three-dimensional imaging through diffusers.}
\label{fig_setup}
\end{figure}

The optical setup in the proposed method is shown in Fig.~\ref{fig_setup}.
A three-dimensional object~$o$ is illuminated with spatially incoherent illumination through a first diffuser, and the light scattered by a second diffuser is captured by a lensless image sensor as $i$.
The relationship between the object and the captured speckle is three-dimensionally shift-invariant, based on the three-dimensional memory effect~\cite{bib_singh2017Scatter}.
Then, the imaging process is written with a scattering impulse response~$h$ as
\begin{equation}
i(\bm{r}_{i})=\int h(\bm{r}_{i}-\bm{r}_{o})o(\bm{r}_{o})\text{d}^{3}{r}_{o},
\label{eq_model}
\end{equation}
where $\bm{r}_i=(x_i,y_i,z_i)$ are the spatial coordinates in the object space, and $\bm{r}_o=(x_o,y_o,z_o)$ are the spatial coordinates in the sensor space, respectively.
The $x$- and $y$-axes are lateral to the image sensor, and the $z$-axis is axial to the image sensor.
The origin of these coordinates is the center of the second diffuser.
The impulse response~$h$ is laterally random and axially scaled with a scaling factor~$s$, which is written as
\begin{equation}
s=(z_o+z_i)/z_o.
\label{eq_mag}
\end{equation}

In our previous work on multi-shot three-dimensional imaging through scattering media~\cite{bib_Okamoto2019Scatter}, the image sensor is axially scanned to observe the three-dimensional speckle distribution~$i$, which is three-dimensionally autocorrelated to remove the impact of the scattering process~$h$ as
\begin{align}
i\star_\text{3D} i&=(h\otimes_\text{3D} o)\star_\text{3D}(h\otimes_\text{3D} o)
\label{eq_corr_1}\\
&=(h\star_\text{3D} h)\otimes_\text{3D}(o\star_\text{3D} o)
\label{eq_corr_2}\\
&\approx o\star_\text{3D} o
\label{eq_corr_3}\\
&=\mathcal{F}_\text{3D}^{-1}[|O|^2],
\label{eq_fourier}
\end{align}
where $\star_\text{3D}$ is the three-dimensional correlation, $\otimes_\text{3D}$ is the three-dimensional convolution,  $\mathcal{F}_\text{3D}$ denotes the three-dimensional Fourier transform, and $O$ is the three-dimensional transform of $o$.
Here, $h\star_\text{3D} h$ is approximated by the delta function due to the laterally random and axially scaled distribution of $h$~\cite{bib_bertolotti2012Scatter, bib_katz2014Speckle}.
The object signal~$o$ is recovered by a phase retrieval process for $|O|^2$~\cite{bib_fienup1982Phase}.

\begin{figure}[]
\begin{center}
		\includegraphics[scale=0.7]{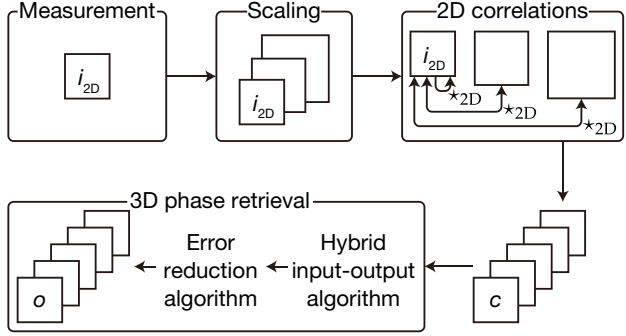}
\end{center}
\caption{Reconstruction process of single-shot three-dimensional imaging by speckle correlation.}
\label{fig_process}
\end{figure}

In this study, as shown in Fig.~\ref{fig_process}, a three-dimensional object is tomographically reconstructed from a single captured speckle image~$i_\text{2D}$, which contains differently scaled impulse responses~$h$ depending on the object distances~$z_o$, using the optical setup in Fig.~\ref{fig_setup}.
The captured speckle image is computationally and laterally scaled multiple times, instead of multiple measurements made while axially scanning the image sensor as in the previous work~\cite{bib_Okamoto2019Scatter}.
This computational scaling process mimics the optical scaling process in Eq.~(\ref{eq_mag}) by axial scanning, where the scaling factor is monotonically increased or decreased.
This process is exploited for searching relative scales of the impulse responses on the original speckle image through the following correlation process, where high correlations appear if the original and scaled impulse responses are coincident, and vice versa.

The scaled speckle images are computationally and laterally correlated as follows:
\begin{equation}
{c^k=}
\left\{
\begin{array}{ll}
      i_\text{2D}^0\star_\text{2D}i_\text{2D}^k,&\text{for}~~k\geq 0\\
  i_\text{2D}^{-k}\star_\text{2D}i_\text{2D}^0,&\text{for}~~k< 0
\end{array}
\right.
\label{corr_2d}
\end{equation}
where $\star_\text{2D}$ is the two-dimensional correlation, the superscript $k=-M+1, -M+2, \cdots, M-2$ denotes the index for the lateral correlations, $M$ is the number of scaled images,
$c^k$ is the $k$-th lateral correlation result, $i_\text{2D}^m$ is the $m$-th scaled speckle image~($i_\text{2D}^0=i_\text{2D}$), and the superscript $m=0, 1, \cdots, M-1$ denotes the index of the scaling factor.
In this study, $o\star_\text{3D} o$ in Eq.~(\ref{eq_corr_3}) is approximated by $c^k$.
The object~$o$ is three-dimensionally reconstructed from the correlation result~$c^k$ based on three-dimensional phase retrieval.

\begin{figure}[]
\begin{center}
		\includegraphics[scale=0.7]{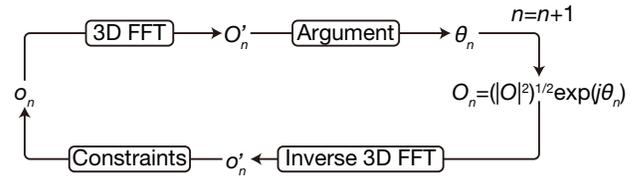}
\end{center}
\caption{Loop in the error reduction algorithm and the hybrid input-output algorithm.}
\label{fig_phase}
\end{figure}

The phase retrieval process in this study follows the previous work on speckle-correlation-based imaging through scattering media~\cite{bib_bertolotti2012Scatter, bib_katz2014Speckle, bib_Okamoto2019Scatter}.
Here two algorithms, which are called the error reduction algorithm and the hybrid input-output algorithm~\cite{bib_fienup1982Phase}, are sequentially performed, as shown in Fig.~\ref{fig_process}.
The second algorithm is a modified version of the first one, and they use the loop in the iterative processes shown in Fig.~\ref{fig_phase}.
The difference between them is the process of constraints.
First, the common aspects are described, and then the differences are explained.

The iterative process of the error reduction algorithm and the hybrid input-output algorithm is shown in Fig.~\ref{fig_phase} and is described as follows:
\begin{enumerate}
\item
The object's Fourier spectrum is initially given with a three-dimensional random phase~$\theta_n$ by $O_n=(|O|^2)^{(1/2)}\exp(j\theta_n)$, where $j$ is the imaginary unit, $|O|^2$ is the three-dimensional Fourier transform of $c^k$ in Eq.~(\ref{corr_2d}), and the subscript~$n$ is the counter of the iteration, which is set to one.
\item
$O_n$ is three-dimensionally inverse Fourier transformed, and the result is set as the intermediately estimated object~$o'_n$.
\item
$o'_n$ is rectified with some constraints, which are described in the following paragraphs, and the result is the estimated object~$o_n$ at the $n$-th iteration. 
\item
$o_n$ is three-dimensionally Fourier transformed, and the result is set as the intermediately estimated object's Fourier spectrum~$O'_n$.
\item
The argument of $O'_n$ is extracted, and it is used as a replacement of the argument~$\theta_n$ of the estimated object's Fourier spectrum~$O_n$, where the counter~$n$ is incremented by one.
\end{enumerate}
Steps~2--5 are iterated.

The rectifying process with constraints at Step~3 is the difference between the error reduction algorithm and the hybrid input-output algorithm~\cite{bib_fienup1982Phase}.
In the error reduction algorithm, the intermediately estimated object~$o'_n$ at the $n$-th iteration is updated with the following rule:
\begin{equation}
o_n(\bm{r}_o)=
\left\{
\begin{array}{ll}
      o'_n(\bm{r}_o),&\text{for}~~\bm{r}_o\notin\Gamma\\
      0,&\text{for}~~\bm{r}_o\in\Gamma
\end{array}
\right.
\label{eq_er}
\end{equation}
where $\Gamma$ is the set of all spatial positions~$\bm{r}_o$ which violate the constraints.
In the hybrid input-output algorithm, the updating rule is as follows:
\begin{equation}
o_n(\bm{r}_o)=
\left\{
\begin{array}{ll}
      o'_n(\bm{r}_o),&\text{for}~~\bm{r}_o\notin\Gamma\\
      o_{n-1}(\bm{r}_o)-\beta o'_n(\bm{r}_o),&\text{for}~~\bm{r}_o\in\Gamma
\end{array}
\right.
\label{eq_hio}
\end{equation}
where $\beta$ is a feedback parameter.

In the reconstruction according to the proposed scheme, as shown in Fig.~\ref{fig_process}, the hybrid input-output algorithm is performed first.
In this algorithm, the feedback parameter~$\beta$ is decreased from 2.0 to 0.0 in intervals of 0.05, and the loop of Fig.~\ref{fig_phase} is iterated ten times for each $\beta$.
Then, the error reduction algorithm is performed by using the result of the hybrid input-output algorithm as the initial estimate at Step~1, and  the loop of Fig.~\ref{fig_phase} is iterated five hundred times.
The constraints used here are realness, non-negativity, and the range of pixel intensities.
Realness and non-negativity are introduced by using spatially incoherent illumination, such as a light emitting diode~(LED) or fluorescence~\cite{bib_bertolotti2012Scatter, bib_katz2014Speckle}.
The range of pixel intensities suppresses some reconstruction artifacts~\cite{bib_Okamoto2019Scatter}.
This phase retrieval has trivial ambiguities of the spatial shift and the conjugate inversion, which have been studied in the literature on phase retrieval~\cite{bib_miao1998Phase}.

\section{Experimental demonstration}

\begin{figure}[]
\begin{center}
		\subfigure[]{\label{fig_exp_obj}\includegraphics[height=2.1cm]{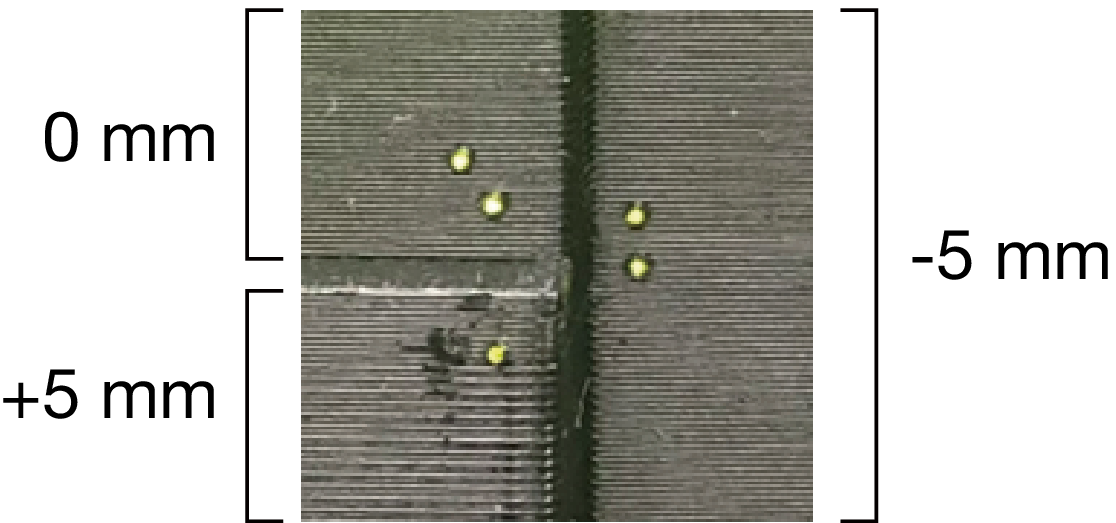}}
		\hspace{0.5cm}
		\subfigure[]{\label{fig_exp_cap}\includegraphics[width=1.7cm]{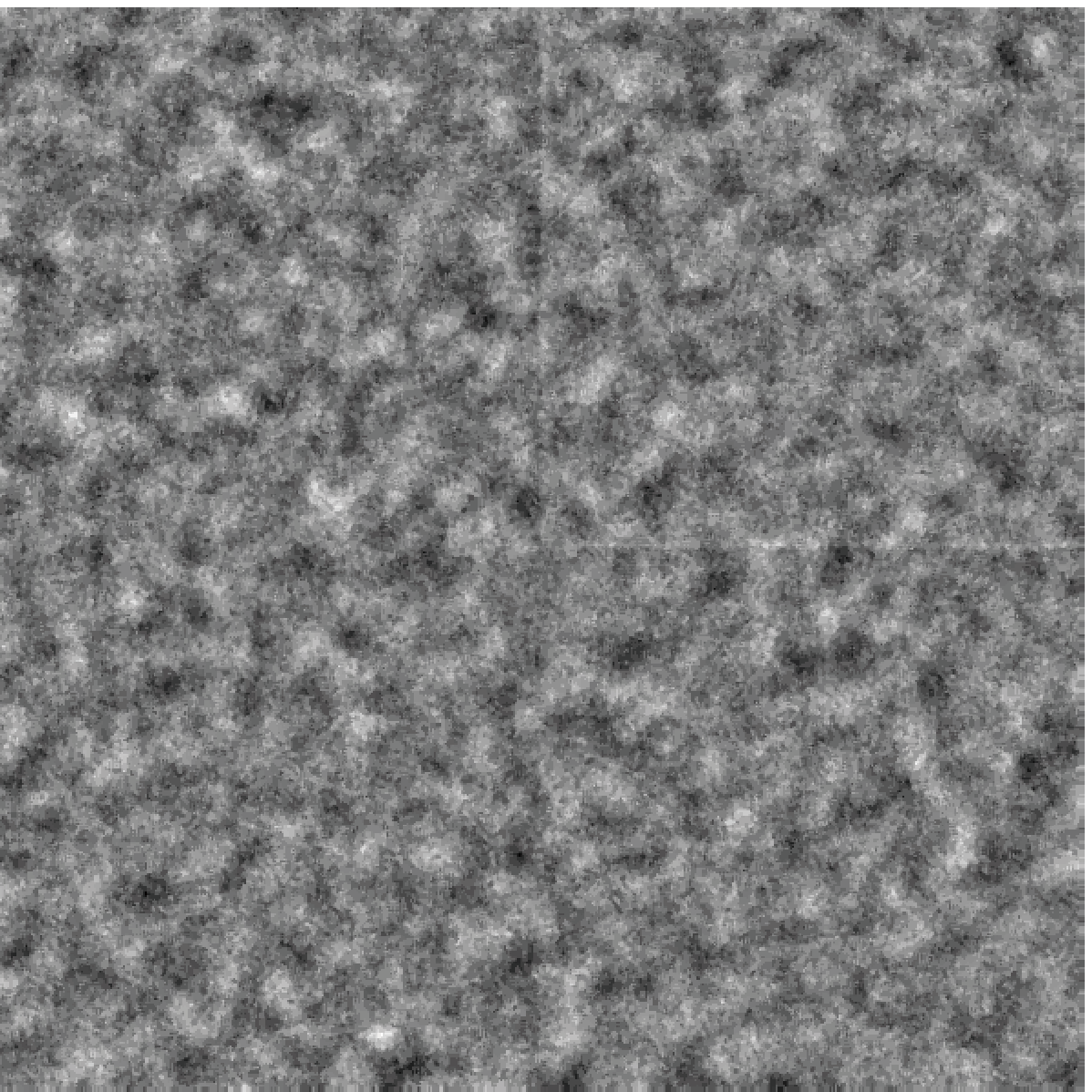}}
		\subfigure[]{\label{fig_exp_corr}\includegraphics[width=\linewidth]{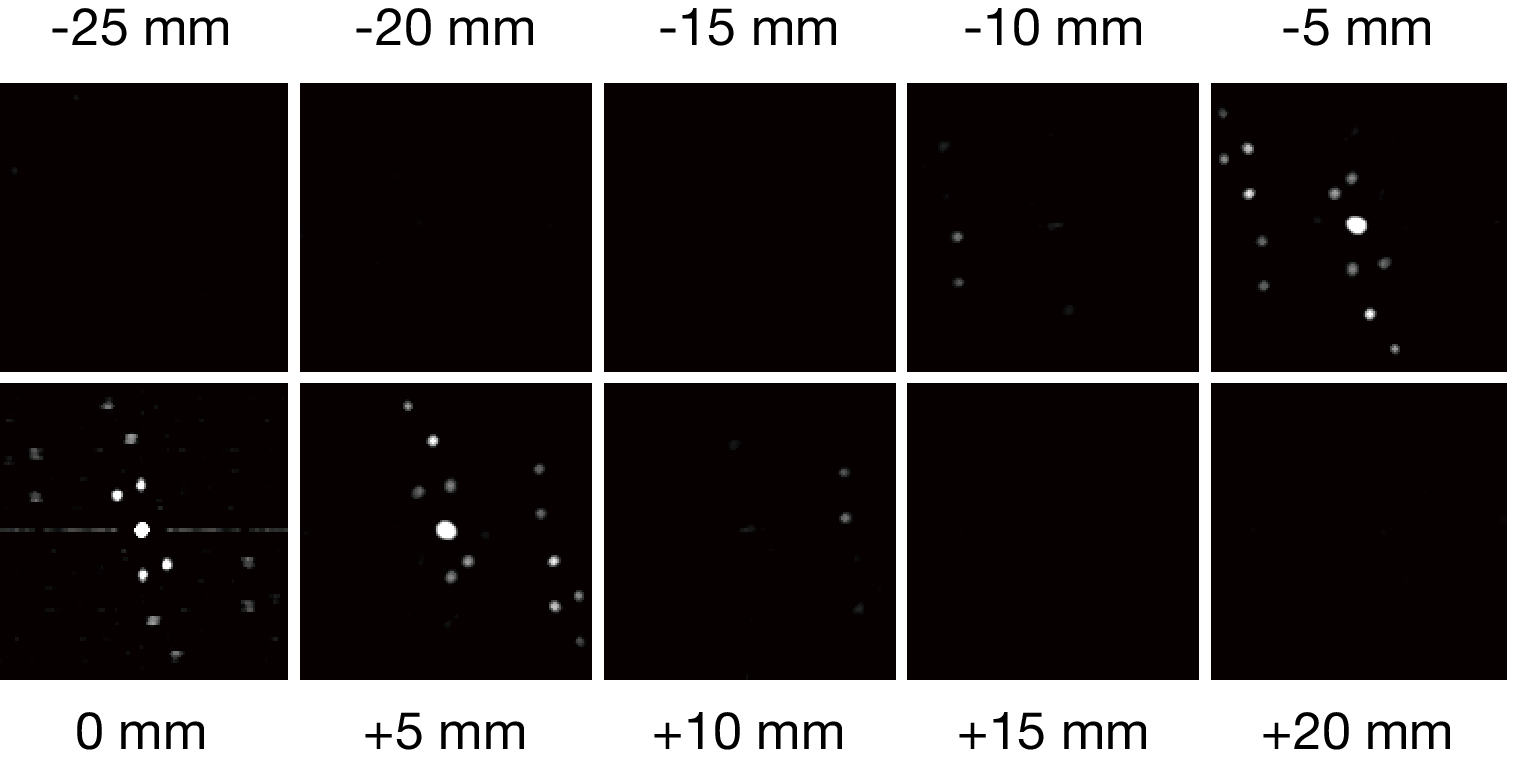}}
		\subfigure[]{\label{fig_exp_recon}\includegraphics[width=\linewidth]{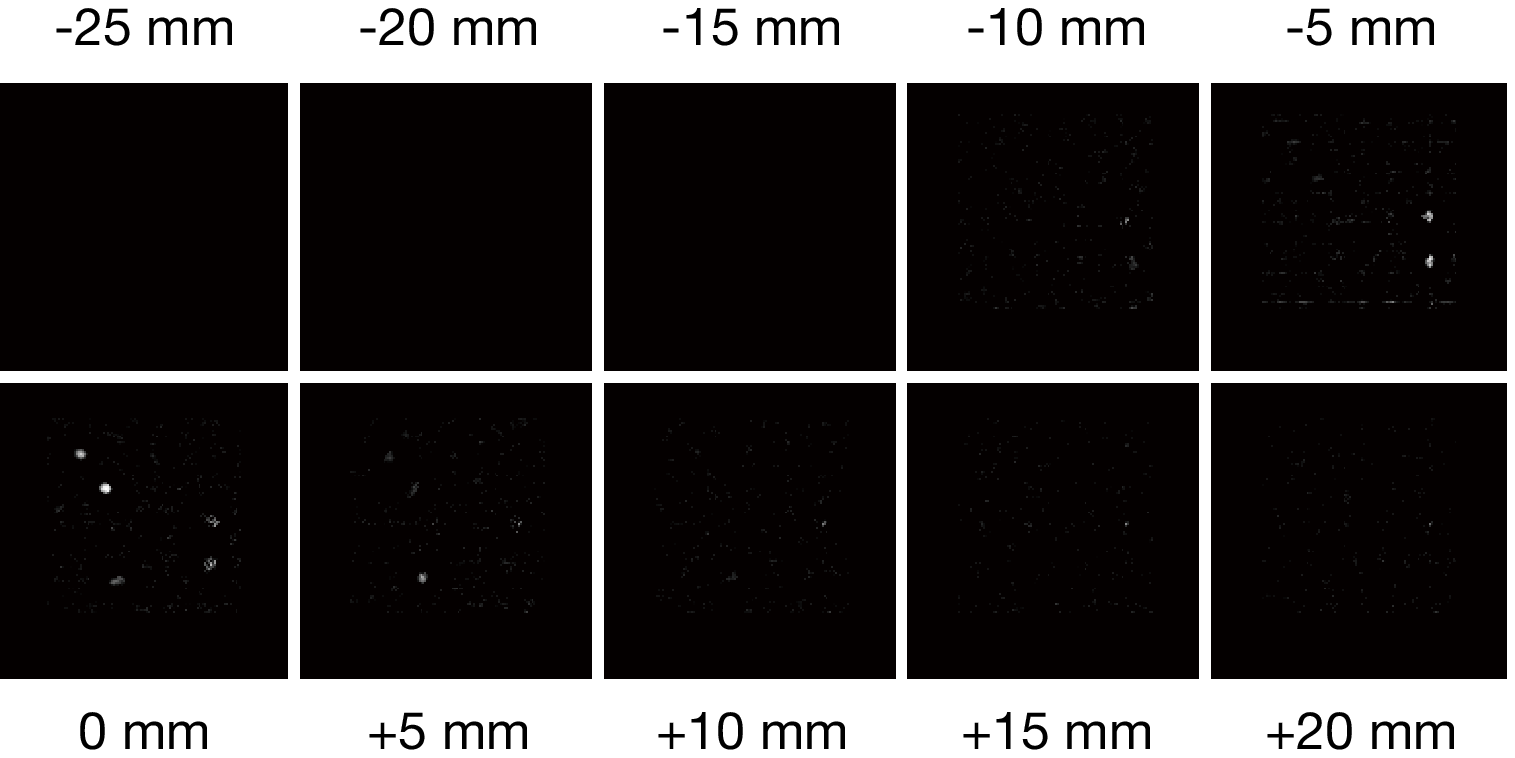}}
		\subfigure[]{\label{fig_exp_recon_multi}\includegraphics[width=\linewidth]{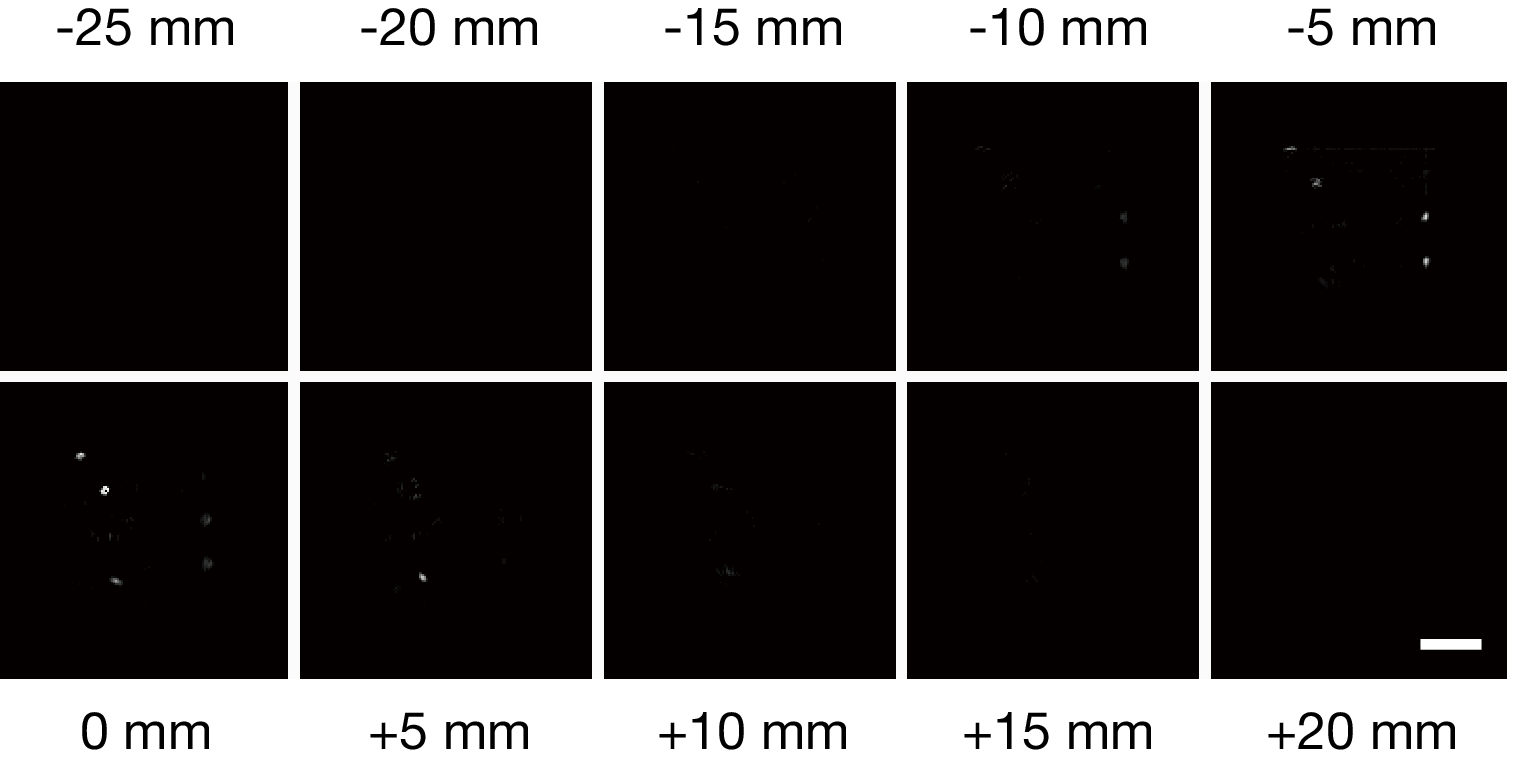}}
\end{center}
\caption{Experimental results.
\subref{fig_exp_obj}~The object.
\subref{fig_exp_cap}~The captured speckle image.
\subref{fig_exp_corr}~The correlation result.
\subref{fig_exp_recon}~The three-dimensional phase retrieval result.
\subref{fig_exp_recon_multi}~The three-dimensional phase retrieval result with the multi-shot approach in Reference~\cite{bib_Okamoto2019Scatter}, where the scale bar is 2~mm at the object plane.
}
\label{fig_exp}
\end{figure}

The proposed method was demonstrated with three-dimensionally arranged point sources fabricated by a 3D printer~(M2030TP manufactured by L-DEVO).
The object had three levels, where the step height was 0.5~cm, and holes with a diameter of 0.4~mm were located at different positions on each level, as shown in Fig.~\ref{fig_exp_obj}.
One hole was made on the front level, two diagonally arranged holes were made on the middle level, and two vertically arranged holes were made on the back level, respectively.
As shown in Fig.~\ref{fig_setup}, the object was located between two diffusers and it was illuminated with an incoherent LED~(M565L3 manufactured by Thorlabs, nominal wavelength:~565~nm, full width at half maximum of spectrum:~103~nm) through a bandpass filter~(578NM X 16NM 25MM manufactured by Edmund Optics, central wavelength~578~nm, full width at half maximum of spectrum:~22~nm) and a diffuser~(LSD5PC10-5 manufactured by Luminit).
The distance between the first diffuser and the object was 75~mm.
Light passing through object was scattered by another diffuser~(LSD20PC10-5 manufactured by Luminit) and captured by a monochrome image sensor~(hr29050MFLGEA manufactured by SVS-Vistek, pixel count:~$4384\times 6576$, pixel pitch:~$5.5\times 5.5$~\textmu m) without any imaging optics.
The distance~($z_o$) between the object and the second diffuser was 92~mm, and the distance~($z_i$) between the second diffuser and the image sensor was 25~mm, as shown in Fig.~\ref{fig_setup}.

The captured speckle image is shown in Fig.~\ref{fig_exp_cap}, where the central $600\times 600$~pixel area was clipped for visualization purposes.
After background compensation, the captured image was scaled with a computational scaling factor~$s_\text{com}^m$ based on Eq.~(\ref{eq_mag}) as follows:
\begin{equation}
s_\text{com}^m=\frac{(z_o-m\Delta_{z_o}+z_i)/(z_o-m\Delta_{z_o})}{(z_o+z_i)/z_o},
\label{eq_comp_mag}
\end{equation}
where $\Delta_{z_o}$ is the axial resolution of the object space, which was set to 0.5~mm, corresponding to the step height of the object.
The number of scaled speckle images~($M$) was set to six.
The scaled speckle images were laterally correlated as in Eq.~(\ref{corr_2d}).
The central $800\times 800$ pixel areas of the correlations were clipped and laterally down-sampled by a factor of four to reduce the computational cost in the next phase retrieval process.
Biases of the correlations were equalized with the average values outside the central areas.
The result of this correlation process is shown in Fig.~\ref{fig_exp_corr}, where the central $150\times 150$ pixel areas were clipped, and contrast enhancement was applied for visualization purposes.

The three-dimensional phase retrieval process was applied to the correlations, and the results are shown in Fig.~\ref{fig_exp_recon}, where the ambiguities of the spatial shift and the conjugate inversion were manually compensated, and the central $150\times 150$~pixel areas were clipped for visualization purposes.
The holes were three-dimensionally recovered.
The reconstruction result of the multi-shot approach with axial scanning of the image sensor in our previous work~\cite{bib_Okamoto2019Scatter} is shown in Fig.~\ref{fig_exp_recon_multi}.
These results were comparable, although some interference from other planes in the single-shot approach was stronger than that in the multi-shot approach.

An issue with the single-shot approach compared with the multi-shot one is distortion of the autocorrelation through the scaling process.
The distortion should be smaller than the lateral resolution of speckle correlations.
The limitation of the scaling factor in Eq.~(\ref{eq_comp_mag}) is calculated as
\begin{equation}
|s_\text{com}^{M-1}-1|d\leq\delta,
\label{eq_lim}
\end{equation}
where $d$ is the object size on the image sensor and is estimated as half of the correlative area, and $\delta$ is the resolution of speckle correlations and is $\sqrt{2}$-times larger than the grain size of the speckles, assuming a Gaussian distribution.
In the case of the experiment, $d=300~\text{pixels}$ and $\delta=\sqrt{2}\times17~\text{pixels}$~\cite{bib_Okamoto2019Scatter}, so the above limitation was satisfied.

\section{Conclusion}

We proposed single-shot three-dimensional imaging through scattering media based on speckle correlation.
An object is three-dimensionally reconstructed from a single speckle image with a scaling process, a correlation process, and a phase retrieval process.
The proposed method was experimentally demonstrated with three-dimensionally arranged point sources between diffusers.
The result was comparable to the multi-shot case reported previously.

Our method simplifies the optical setup for three-dimensional imaging through scattering media.
It is useful in a wide range of modalities, including lensless imaging, microscope imaging, and telescope imaging, and applications such as looking around corners and reflection-mode scattered imaging.
Also, this method may provide interesting insights for three-dimensional imaging without the need to calibrate optical hardware or probe optical phenomena before an imaging stage.

\section*{Funding}
This work was supported by JSPS KAKENHI Grant Numbers~JP17H02799 and JP17K00233, JST PRESTO Grant Number~JPMJPR17PB.


\begin{thebibliography}{10}
\newcommand{\enquote}[1]{``#1''}

\bibitem{bib_merali2015Scatter}
Z.~Merali, \enquote{{Optics: Super vision},} {\protect\JournalTitle{Nature}}
  \textbf{518}, 158--160 (2015).

\bibitem{bib_gigan2017Scatter}
S.~Gigan, \enquote{{Optical microscopy aims deep},}
  {\protect\JournalTitle{Nature Photonics}} \textbf{11}, 14--16 (2017).

\bibitem{bib_Marx2017AO}
V.~Marx, \enquote{{Microscopy: Hello, adaptive optics},}
  {\protect\JournalTitle{Nature Methods}} \textbf{14}, 1133--1136 (2017).

\bibitem{bib_Watnik2018Wavefront}
A.~T. Watnik and D.~F. Gardner, \enquote{{Wavefront sensing in deep
  turbulence},} {\protect\JournalTitle{Optics and Photonics News}} \textbf{29},
  38--45 (2018).

\bibitem{bib_mosk2012Scatter}
A.~P. Mosk, A.~Lagendijk, G.~Lerosey, and M.~Fink, \enquote{{Controlling waves
  in space and time for imaging and focusing in complex media},}
  {\protect\JournalTitle{Nat. Photonics}} \textbf{6}, 283--292 (2012).

\bibitem{bib_horstmeyer2015Scatter}
R.~Horstmeyer, H.~Ruan, and C.~Yang, \enquote{{Guidestar-assisted
  wavefront-shaping methods for focusing light into biological tissue},}
  {\protect\JournalTitle{Nat. Photonics}} \textbf{9}, 563--571 (2015).

\bibitem{bib_vellekoop2007Focusing}
I.~M. Vellekoop and A.~P. Mosk, \enquote{{Focusing coherent light through
  opaque strongly scattering media},} {\protect\JournalTitle{Opt. Lett.}}
  \textbf{32}, 2309--2311 (2007).

\bibitem{bib_aulbach2011Scatter}
J.~Aulbach, B.~Gjonaj, P.~M. Johnson, A.~P. Mosk, and A.~Lagendijk,
  \enquote{{Control of light transmission through opaque scattering media in
  space and time},} {\protect\JournalTitle{Phys. Rev. Lett.}} \textbf{106},
  103901 (2011).

\bibitem{bib_katz2011Scatter}
O.~Katz, E.~Small, Y.~Bromberg, and Y.~Silberberg, \enquote{{Focusing and
  compression of ultrashort pulses through scattering media},}
  {\protect\JournalTitle{Nat. Photonics}} \textbf{5}, 372--377 (2011).

\bibitem{bib_katz2012Scatter}
O.~Katz, E.~Small, and Y.~Silberberg, \enquote{{Looking around corners and
  through thin turbid layers in real time with scattered incoherent light},}
  {\protect\JournalTitle{Nat. Photonics}} \textbf{6}, 549--553 (2012).

\bibitem{bib_vellekoop2010Foucising}
I.~M. Vellekoop, A.~Lagendijk, and A.~P. Mosk, \enquote{{Exploiting disorder
  for perfect focusing},} {\protect\JournalTitle{Nat. Photonics}} \textbf{4},
  320--322 (2010).

\bibitem{bib_park2013Nanoparticle}
J.-H. Park, C.~Park, H.~Yu, J.~Park, S.~Han, J.~Shin, S.~H. Ko, K.~T. Nam,
  Y.-H. Cho, and Y.~Park, \enquote{{Subwavelength light focusing using random
  nanoparticles},} {\protect\JournalTitle{Nat. Photonics}} \textbf{7}, 454--458
  (2013).

\bibitem{bib_conkey2012Scatter}
D.~B. Conkey, A.~N. Brown, A.~M. Caravaca-Aguirre, and R.~Piestun,
  \enquote{{Genetic algorithm optimization for focusing through turbid media in
  noisy environments},} {\protect\JournalTitle{Opt. Express}} \textbf{20},
  4840--4849 (2012).

\bibitem{bib_Conkey2012ColorScatter}
D.~B. Conkey and R.~Piestun, \enquote{{Color image projection through a
  strongly scattering wall},} {\protect\JournalTitle{Opt. Express}}
  \textbf{20}, 27312--27318 (2012).

\bibitem{bib_vellekoop2015Scatter}
I.~M. Vellekoop, \enquote{{Feedback-based wavefront shaping},}
  {\protect\JournalTitle{Opt. Express}} \textbf{23}, 12189--12206 (2015).

\bibitem{bib_yaqoob2008Scatter}
Z.~Yaqoob, D.~Psaltis, M.~S. Feld, and C.~Yang, \enquote{{Optical phase
  conjugation for turbidity suppression in biological samples},}
  {\protect\JournalTitle{Nat. Photonics}} \textbf{2}, 110--115 (2008).

\bibitem{bib_popoff2010Scatter}
S.~M. Popoff, G.~Lerosey, R.~Carminati, M.~Fink, A.~C. Boccara, and S.~Gigan,
  \enquote{{Measuring the transmission matrix in optics: an approach to the
  study and control of light propagation in disordered media},}
  {\protect\JournalTitle{Phys. Rev. Lett.}} \textbf{104}, 100601 (2010).

\bibitem{bib_popoff2010Opaque}
S.~Popoff, G.~Lerosey, M.~Fink, A.~C. Boccara, and S.~Gigan, \enquote{{Image
  transmission through an opaque material},} {\protect\JournalTitle{Nature
  Communications}} \textbf{1}, 1--5 (2010).

\bibitem{bib_popoff2011Scatter}
S.~M. Popoff, G.~Lerosey, M.~Fink, A.~C. Boccara, and S.~Gigan,
  \enquote{{Controlling light through optical disordered media: transmission
  matrix approach},} {\protect\JournalTitle{New Journal of Physics}}
  \textbf{13}, 123021 (2011).

\bibitem{bib_Lai2012Scatter}
X.~Xu, H.~Liu, and L.~V. Wang, \enquote{{Time-reversed ultrasonically encoded
  optical focusing into scattering media},} {\protect\JournalTitle{Nature
  Photonics}} \textbf{5}, 154--157 (2011).

\bibitem{bib_chaigne2013Scatter}
T.~Chaigne, O.~Katz, A.~C. Boccara, M.~Fink, E.~Bossy, and S.~Gigan,
  \enquote{{Controlling light in scattering media non-invasively using the
  photoacoustic transmission matrix},} {\protect\JournalTitle{Nat. Photonics}}
  \textbf{8}, 58--64 (2013).

\bibitem{bib_benjamin2013Scatter}
B.~Judkewitz, Y.~M. Wang, R.~Horstmeyer, A.~Mathy, and C.~Yang,
  \enquote{{Speckle-scale focusing in the diffusive regime with time reversal
  of variance-encoded light (TROVE)},} {\protect\JournalTitle{Nature
  Photonics}} \textbf{7}, 300--305 (2013).

\bibitem{bib_liutkus2014CS}
A.~Liutkus, D.~Martina, S.~Popoff, G.~Chardon, O.~Katz, G.~Lerosey, S.~Gigan,
  L.~Daudet, and I.~Carron, \enquote{{Imaging with nature: compressive imaging
  using a multiply scattering medium},} {\protect\JournalTitle{Sci. Rep.}}
  \textbf{4}, 5552 (2014).

\bibitem{bib_mounaix2016Scatter}
M.~Mounaix, H.~Defienne, and S.~Gigan, \enquote{{Deterministic light focusing
  in space and time through multiple scattering media with a time-resolved
  transmission matrix approach},} {\protect\JournalTitle{Phys. Rev. A}}
  \textbf{94}, 41802 (2016).

\bibitem{bib_horisaki2016Scatter}
R.~Horisaki, R.~Takagi, and J.~Tanida, \enquote{{Learning-based imaging through
  scattering media},} {\protect\JournalTitle{Opt. Express}} \textbf{24},
  13738--13743 (2016).

\bibitem{bib_edrei2016ScatterDeconv}
E.~Edrei and G.~Scarcelli, \enquote{{Memory-effect based deconvolution
  microscopy for super-resolution imaging through scattering media},}
  {\protect\JournalTitle{Scientific Reports}} \textbf{6}, 33558 (2016).

\bibitem{bib_Horisaki2017Scatter}
R.~Horisaki, R.~Takagi, and J.~Tanida, \enquote{{Learning-based focusing
  through scattering media},} {\protect\JournalTitle{Appl. Opt.}} \textbf{56},
  4358--4362 (2017).

\bibitem{bib_Horisaki2017SR}
R.~Horisaki, R.~Takagi, and J.~Tanida, \enquote{{Learning-based single-shot
  superresolution in diffractive imaging},} {\protect\JournalTitle{Appl. Opt.}}
  \textbf{56}, 8896--8901 (2017).

\bibitem{bib_Singh2017ScatterPlate}
A.~K. Singh, G.~Pedrini, M.~Takeda, and W.~Osten, \enquote{{Scatter-plate
  microscope for lensless microscopy with diffraction limited resolution},}
  {\protect\JournalTitle{Scientific Reports}} \textbf{7}, 10687 (2017).

\bibitem{bib_Antipa2018Diffuser}
N.~Antipa, G.~Kuo, R.~Heckel, B.~Mildenhall, E.~Bostan, R.~Ng, and L.~Waller,
  \enquote{{DiffuserCam: lensless single-exposure 3D imaging},}
  {\protect\JournalTitle{Optica}} \textbf{5}, 1--9 (2018).

\bibitem{bib_Feng1988Scatter}
S.~Feng, C.~Kane, P.~A. Lee, and A.~D. Stone, \enquote{{Correlations and
  fluctuations of coherent wave transmission through disordered media},}
  {\protect\JournalTitle{Phys. Rev. Lett.}} \textbf{61}, 834--837 (1988).

\bibitem{bib_Freund1990Scatter}
I.~Freund, \enquote{{Looking through walls and around corners},}
  {\protect\JournalTitle{Physica A: Statistical Mechanics and its
  Applications}} \textbf{168}, 49--65 (1990).

\bibitem{bib_bertolotti2012Scatter}
J.~Bertolotti, E.~G. van Putten, C.~Blum, A.~Lagendijk, W.~L. Vos, and A.~P.
  Mosk, \enquote{{Non-invasive imaging through opaque scattering layers},}
  {\protect\JournalTitle{Nature}} \textbf{491}, 232--234 (2012).

\bibitem{bib_katz2014Speckle}
O.~Katz, P.~Heidmann, M.~Fink, and S.~Gigan, \enquote{{Non-invasive single-shot
  imaging through scattering layers and around corners via speckle
  correlations},} {\protect\JournalTitle{Nat. Photonics}} \textbf{8}, 784--790
  (2014).

\bibitem{bib_singh2014Diffuse}
A.~K. Singh, D.~N. Naik, G.~Pedrini, M.~Takeda, and W.~Osten, \enquote{{Looking
  through a diffuser and around an opaque surface: A holographic approach},}
  {\protect\JournalTitle{Opt. Express}} \textbf{22}, 7694--7701 (2014).

\bibitem{bib_schott2015Scatter}
S.~Schott, J.~Bertolotti, J.-F. L{\'{e}}ger, L.~Bourdieu, and S.~Gigan,
  \enquote{{Characterization of the angular memory effect of scattered light in
  biological tissues},} {\protect\JournalTitle{Opt. Express}} \textbf{23},
  13505--13516 (2015).

\bibitem{bib_edrei2016Scatter}
E.~Edrei and G.~Scarcelli, \enquote{{Optical imaging through dynamic turbid
  media using the Fourier-domain shower-curtain effect},}
  {\protect\JournalTitle{Optica}} \textbf{3}, 71--74 (2016).

\bibitem{bib_li2017ScatterEncryption}
G.~Li, W.~Yang, D.~Li, and G.~Situ, \enquote{{Cyphertext-only attack on the
  double random-phase encryption: Experimental demonstration},}
  {\protect\JournalTitle{Opt. Express}} \textbf{25}, 8690--8697 (2017).

\bibitem{bib_Yang2018Scatter}
W.~Yang, G.~Li, and G.~Situ, \enquote{{Imaging through scattering media with
  the auxiliary of a known reference object},}
  {\protect\JournalTitle{Scientific Reports}} \textbf{8}, 9614 (2018).

\bibitem{bib_takasaki2014Scatter}
K.~T. Takasaki and J.~W. Fleischer, \enquote{{Phase-space measurement for
  depth-resolved memory-effect imaging},} {\protect\JournalTitle{Opt. Express}}
  \textbf{22}, 31426--31433 (2014).

\bibitem{bib_singh2017Scatter}
A.~K. Singh, D.~N. Naik, G.~Pedrini, M.~Takeda, and W.~Osten,
  \enquote{{Exploiting scattering media for exploring 3D objects},}
  {\protect\JournalTitle{Light: Science {\&} Applications}} \textbf{6}, e16219
  (2016).

\bibitem{bib_Shi2017Scatter}
Y.~Shi, Y.~Liu, J.~Wang, and T.~Wu, \enquote{{Non-invasive depth-resolved
  imaging through scattering layers via speckle correlations and parallax},}
  {\protect\JournalTitle{Applied Physics Letters}} \textbf{110}, 231101 (2017).

\bibitem{bib_Okamoto2019Scatter}
Y.~Okamoto, R.~Horisaki, and J.~Tanida, \enquote{{Noninvasive three-dimensional
  imaging through scattering media by three-dimensional speckle correlation},}
  {\protect\JournalTitle{Opt. Lett.}} \textbf{44}, 2526--2529 (2019).

\bibitem{bib_fienup1982Phase}
J.~R. Fienup, \enquote{{Phase retrieval algorithms: a comparison},}
  {\protect\JournalTitle{Appl. Opt.}} \textbf{21}, 2758--2769 (1982).

\bibitem{bib_miao1998Phase}
J.~Miao, D.~Sayre, and H.~N. Chapman, \enquote{{Phase retrieval from the
  magnitude of the Fourier transforms of nonperiodic objects},}
  {\protect\JournalTitle{J. Opt. Soc. Am. A}} \textbf{15}, 1662--1669 (1998).

\end{thebibliography}

\end{document}